\documentclass[12pt]{iopart}
\usepackage{iopams}
\usepackage{graphicx}
\newcommand{\ce}{CeCu$_2$Si$_2$}

\newcommand{\cps}{CePd$_2$Si$_2$}
\newcommand{\tc}{$T_c$}
\newcommand{\tconset}{$T_c^{\rm onset}$}
\newcommand{\sllc}{($\sigma\| c$)}
\newcommand{\sprc}{($\sigma\bot c$)}

\newcommand{\pv}{$P_{\rm v}$}

\begin{document}

\title{Anisotropy, disorder, and superconductivity in \ce\ under high pressure}

\author{A T Holmes$^{1,2}$, D Jaccard$^2$, H S
Jeevan$^3$, C Geibel$^3$ and M Ishikawa$^4$
 }

\address{$^1$ KYOKUGEN, Osaka University, Toyonaka, 560-8531, Japan}
\address{$^2$ DPMC, University of Geneva, 24 Quai Ernest-Ansermet, CH-1211
Geneva 4, Switzerland}
\address{$^3$\ Max-Planck Institute for Chemical Physics of Solids, D-01187 Dresden, Germany}
\address{$^4$ Institute for Solid State Physics, University of Tokyo, Kashiwa, Chiba 277-8581, Japan}
\ead{alex@djebel.mp.es.osaka-u.ac.jp}
\begin{abstract}
Resistivity measurements were carried out up to 8$\:$GPa on single
crystal and polycrystalline samples of \ce\ from differing sources
in the homogeneity range. The anisotropic response to current
direction and small uniaxial stresses was explored, taking
advantage of the quasi-hydrostatic environment of the Bridgman
anvil cell.  It was found that both the superconducting transition
temperature \tc\ and the normal state properties are very
sensitive to uniaxial stress, which leads to a shift of the
valence instability pressure \pv\ and a small but significant
change in \tc\ for different orientations with respect to the
tetragonal $c$-axis. Coexistence of superconductivity and residual
resistivity close to the Ioffe-Regel limit around 5$\:$GPa
provides a compelling argument for the existence of a
valence-fluctuation mediated pairing interaction at high pressure
in \ce.

\end{abstract}

\pacs{74.62.Fj, 74.25.Fy, 74.20.Mn}



\section{Introduction}
It has recently been proposed that a new type of superconductivity
exists in the heavy fermion \ce\ under high
pressure\cite{Miyake99,Onishi00,Holmes04a,Yuan03b,Jaccard04}. In
this compound, the superconducting transition temperature $T_c$ is
enhanced near 3$\:$GPa from its ambient pressure value of 0.7$\:$K
to around 2.5$\:$K. The proposed pairing mechanism at high
pressure is based on the exchange of critical valence fluctuations
close to a (nearly) first order valence transition of the Ce ion
at a pressure $P_{\rm v}\simeq$ 4.5$\:$GPa. At ambient pressure in
contrast, critical magnetic fluctuations are believed to mediate
the superconductivity. The magnetically ordered state, which in
pure \ce\ competes with the superconducting state is thought to
disappear at a magnetic quantum critical point at a small positive
pressure $P_c$ of  approximately 0.1$\:$GPa \cite{Gegenwart98}.
This quantum critical point is masked by the superconducting state
in pure samples, but can be directly observed by substituting Ge
for Si which leads to a suppression of superconductivity
\cite{Yuan03b,Trovarelli97}.

The necessary background to this work can be found in
Ref.~\cite{Holmes04a}, which provides the context for the results
and discussion reported below, and should be read in conjunction
with this paper. The most important point to remember is that
there is a second critical pressure \pv\ where the occupation
number $n_f$ of the electronic $f$-orbitals on the Ce atom changes
abruptly. This can be thought of as analogous to the Ce
$\alpha/\gamma$ transition, though with the critical endpoint at
very low, or perhaps negative temperature (the latter leading to a
crossover rather than a first order transition). The system goes
from a Kondo regime ($n_f\simeq 1$) to that with characteristics
of an intermediate valence state (where $n_f<1$). The enhancement
of \tc\ near \pv, mediated by critical valence fluctuations, was
predicted by an extended Anderson model including an extra Coulomb
repulsion term $U_{cf}$ between the conduction $c$- and
$f$-electrons~\cite{Miyake99,Onishi00}. A series of anomalies in
the normal state electronic properties was also predicted, and
these have been observed to occur around this pressure. They
include a large enhancement of the residual resistivity $\rho_0$,
an enhancement of the Sommerfeld coefficient $\gamma$, and a
resistivity linear in temperature, $(\rho-\rho_0)\propto
T$~\cite{Onishi00,Holmes04a,Miyake02}. This simple model is
surprisingly successful in describing the observed behaviour in
\ce, and can help to provide the basis for a more complete
description of the behaviour of related systems. However, there
are further experimental facts beyond its scope, which may prove
useful guidance towards a more complete understanding of the
microscopic physics involved.

 The superconducting and magnetic properties of
\ce\ at ambient pressure are highly sensitive to small changes in
composition and disorder
\cite{Ishikawa83,Modler95,Steglich96,Louca00}, but there has been
little systematic investigation into similar effects at high
pressure.  The ground state at ambient pressure depends strongly
on the exact stoichiometry of the sample, giving so-called types
A, A/S and S, where the magnetically ordered `A phase' competes
with the superconducting state `S'. Ishikawa \cite{Ishikawa03}
recently proposed a further subdivision of \ce\ properties, into
so-called `low \tc ' and `high \tc ' types, with differing signs
in the pressure dependence of the superconducting transition
temperature \tc. In addition, the isostructural compound \cps\ has
been shown to be extremely sensitive to uniaxial stress under
pressure. Dramatic variations in $T_c$ in \cps\ result from a
change in crystalline orientation with respect to small
non-hydrostatic components in a Bridgman anvil pressure cell with
a quasi-hydrostatic steatite medium \cite{Demuer02}.

With these facts in mind, we wished to systematically explore, via
resistance measurements under pressure, how the electronic
properties around \pv\ depended on the sample, and on the pressure
conditions; especially to see how valence-fluctuation mediated
superconductivity in \ce\ is affected by the presence of disorder
and uniaxial stress. The results previously obtained in a
hydrostatic helium pressure medium \cite{Holmes04a} provide a
baseline for comparison.

\section{Experimental methods}

Resistivity measurements under pressure on a total of six \ce\
samples from two different sources are reported below. There were
two polycrystalline samples, of type `low-\tc ' and `high-\tc ',
respectively labelled \#50 and \#57 and prepared by Ishikawa's
group by a levitation method; the remaining four samples were A/S
type single crystals from the same original source crystal
prepared at the MPI Dresden. The latter crystal were grown in an
Al$_2$O$_3$ crucible by a modified Bridgman technique using Cu
excess as flux medium \cite{Jeevanpc}. X-ray powder diffraction
confirmed the tetragonal ThCr$_2$Si$_2$ structure with lattice
parameters $a = 4.099\:\AA$ and $c = 9.922\:\AA$. Specific heat
measurements display a large mean field like anomaly at $T_A =
0.7\:$K followed by a large peak at $T_c = 0.5\:$K. According to
the strong difference in their magnetic field dependence, these
anomalies can be attributed to the transition into the magnetic
A-Phase and into the superconducting state, respectively. The
large size of the anomalies proves that both transitions are bulk
ones.

Two different pressure runs are reported in this paper, referred
to below as cell \#1 (containing four samples from all sources)
and \#2 (with two A/S samples) using a Bridgman anvil cell
technique with a quasi-hydrostatic steatite solid pressure medium.
Pressure gradients are believed to be less than 5\% of the total
at the highest pressures, determined from the width of the
superconducting transition of Pb manometer. The first cell (\#1)
contained the two polycrystals \#50 and \#57 and two A/S type
single crystals, oriented with the measurement current parallel to
the $c$- and $a$-axes, and labelled `A/S $I\parallel c$' and `A/S
$I\parallel a$' respectively. The current direction was
perpendicular to the axis of the pressure cell, and we believe
though unfortunately are not certain, that the $c$-axis of sample
`A/S $I\parallel a$' was oriented parallel to the cell axis.

\begin{figure}
\begin{center}
\includegraphics[width=.6\columnwidth]{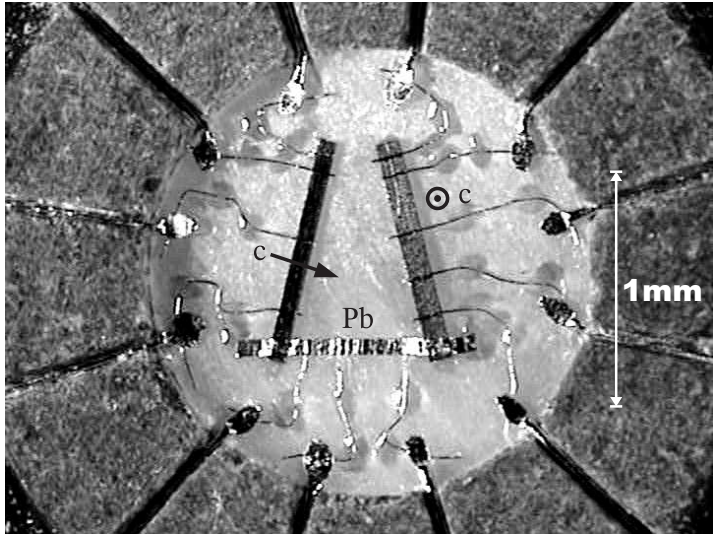}
\end{center}
\caption{\label{fig:strainphoto} Cell \#2, containing two A/S
samples cut from neighbouring positions on the same source
crystal.  They are oriented with their $c$-axes parallel and
perpendicular to the cell axis, and the current flows along the
$a$ direction.}
\end{figure}
Any uniaxial stresses due to an imperfectly hydrostatic pressure
medium are likely to be aligned with the cell axis, and the second
pressure cell (\#2), Fig.~\ref{fig:strainphoto}, was designed to
exploit this systematically. The cell contained two samples, both
cut from neighbouring positions on the same A/S type single
crystal source. They were both oriented so that the current flowed
along the $a$ direction, while the $c$-axis was oriented parallel
and perpendicular to the cell axis. The samples have been labelled
\sllc\ and \sprc\ respectively. Multiple contacts on each sample
enabled the resistivity within different regions of the same
sample to be compared. As they shared their longest side in the
original source crystal, variations in composition along the
length of one sample should be comparable to that between samples.

\section{Results}
The resistivity was first measured at P=0; excellent agreement was
found with previous results. Apart from the superconducting
transition temperature \tc, several features of the normal state
resistivity under pressure will be highlighted. These are the
residual resistivity $\rho_0$ and $A$ coefficient of the
temperature dependence in the power law behaviour
$\rho=\rho_0+AT^n$ of the normal state at low temperature, and the
position of the two crystal-field split resistivity maxima
$T_1^{\rm max}$ and $T_2^{\rm max}$. All of these show distinctive
features at the valence instability pressure \pv.

The resistance has been normalized by the geometric factor of each
sample measured on construction of the cell, which usually gives
an absolute value within 10\% for proper four-point measurements.
Contrary to cell \#2, which provided straightforward results, the
gasket of cell \#1 formed under too high a load, and due to
migration of the contact wires, full four-point resistance
measurements were possible in only one of the samples (\#50). The
resistance measured for the other samples contained an additional
series contribution, of varying size. It would be possible to
subtract this, for example assuming a linear additional term
and/or an adjustment of the geometric factor used to obtain the
absolute resistivity, but the results obtained will then depend on
these assumptions. We should add that obtaining even
semi-quantitative values for the absolute resistivity at such high
pressures is a challenging task, therefore we make no apology for
trying to extract the maximum amount of information from what
might be seen as an imperfect experiment.

\begin{figure}[t]
\begin{center}
\includegraphics[width=0.6\columnwidth]{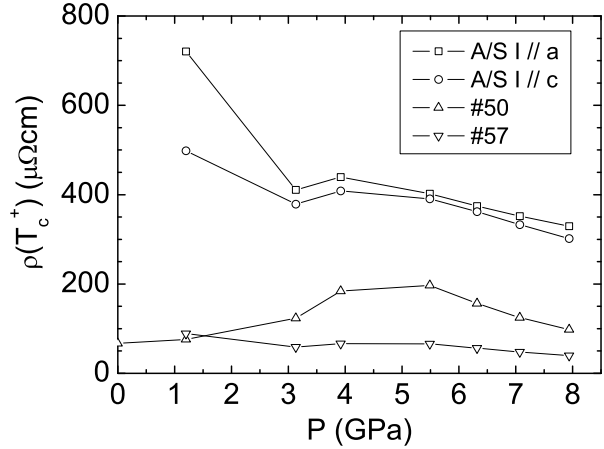}
\end{center}
\caption{\label{fig:rho0} Normal state resistivity at \tc\ in cell
\#1, showing the enhancement of $\rho_0$ at \pv.  Note that only
sample \#50 had full four-point contacts, so the others contain a
monotonically decreasing additional contribution to the
resistance, but the sample contribution is visible in all.}
\end{figure}

In Fig.~\ref{fig:rho0}, one can easily see the enhancement of the
residual resistivity at \pv\ in all samples in cell \#1,
superimposed in three cases on a monotonically decreasing
additional series resistance.  It is clear that the enhancement of
the residual resistivity in sample \#50 (the sample with four
electrical contacts) is much larger than in the others, and
reaches almost 200$\:\mu\Omega$cm close to 5$\:$GPa. This sample
also has a small negative magnetoresistance at 5.5$\:$GPa, around
0.5\% at 8$\:$T and 4.2$\:$K, in contrast to the other samples
which show a positive magnetoresistance. This is typical of \ce\
samples with very high residual resistivities at $P_{\rm v}$, and
perhaps due to the Zeeman shift of the $f$-level $\epsilon_f$,
moving the system away from the valence instability.

\begin{figure}[b]
\begin{center}
\includegraphics[width=0.6\columnwidth]{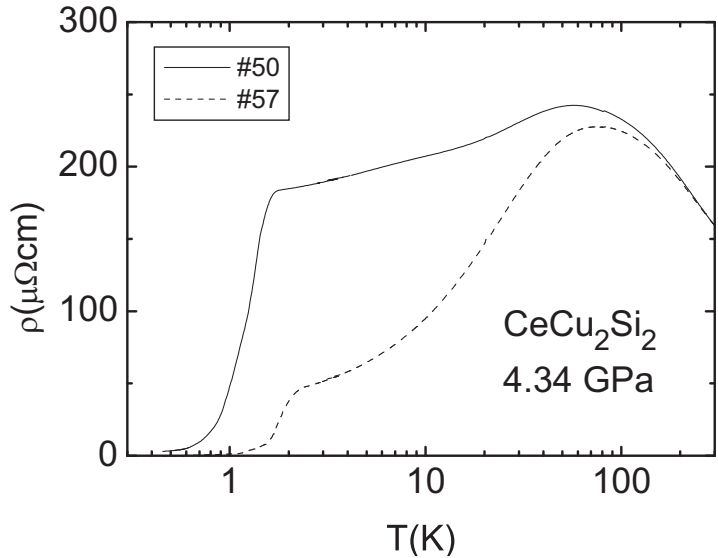}
\end{center}
\caption{\label{fig:5057p4} Resistivity of the `low $T_c$' (\#50,
unadjusted) and `high $T_c$' (\#57, adjusted, see text) samples in
cell \#1 at 4.34 GPa. Note the extremely large residual
resistivity of \#50, yet a nearly complete superconducting
transition is observed.}
\end{figure}
In Fig.~\ref{fig:5057p4}, samples \#50 and \#57 are compared at
4.34$\:$GPa, close to the maximum of $\rho_0$.  To compensate for
experimental difficulties, the resistivity of \#57 has been
adjusted slightly to give zero resistance below \tc, and to match
that of \#50 at room temperature. This was done by subtracting a
small constant contact resistance and slightly adjusting the
geometric factor, each corresponding to about 10\% of the
unadjusted data. The resulting resistivity curve is typical of
\ce\ at this pressure. In sample \#50 however, there is evidently
an enormously enhanced impurity contribution to the total
resistivity, with a negative temperature dependence reminiscent of
a Kondo impurity system, in addition to the usual scattering which
increases up to $T^{\rm max}\sim T_K$. A negative slope becomes
apparent at higher pressure where the $A$ coefficient of the
resistivity has collapsed and the impurity contribution dominates
the resistivity. The two samples regain very similar behaviour on
approaching room temperature. We should add that this sort of
behaviour, where superconductivity coexists with very high
residual resistivity, has also been observed in single crystals
\cite{Jaccard98}. The existence of samples which superconduct at
high pressure despite such enormous residual resistivities, of the
order of the Ioffe-Regel limit (which is around 100$\:\mu\Omega$cm
at ambient pressure), is strong evidence in favour of valence
fluctuation mediated pairing. The common view is that
unconventional superconductivity is incompatible with large
electronic scattering rates. However, in this case both the
superconductivity and the enhanced impurity scattering may share a
common origin, as we believe that valence fluctuations are
responsible for both the pairing interaction and the
renormalization of impurity potentials \cite{Onishi00,Miyake02}.

\begin{figure}
\begin{center}
\includegraphics[width=0.6\columnwidth]{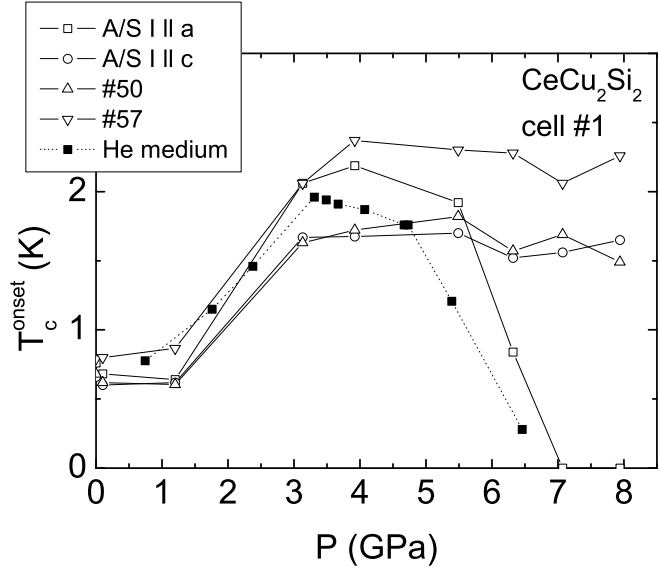}
\end{center}
\caption{\label{fig:Tconset} The superconducting pressure range is
extended considerably in three out of four samples. Only sample
`A/S $I\parallel a$' showed the expected total disappearance of
superconductivity above 7$\:$GPa.}
\end{figure}

Figure \ref{fig:Tconset} shows the onset temperature of the
superconducting transition \tconset\ for the samples in cell \#1.
The most remarkable feature is that in three out of four samples,
the superconducting region extends up to a much higher pressure
than seen in hydrostatic conditions. Only in sample `A/S
$I\parallel a$' was \tconset\ suppressed at high pressures similar
to the result observed in the helium cell. The transitions at the
highest pressures are partial, but there is a clear drop in
resistance below \tconset.  Very broad resistive transitions are
intrinsic to \ce\ at high pressure, and have been identified as
due to filamentary superconductivity rather than to pressure
gradients \cite{Holmes04a}.  The superconducting state at very
high pressure may therefore be solely of a filamentary nature. The
enhanced \tconset\ at very high pressure is a clear indication of
the sensitivity of the high pressure superconducting state to
anisotropy, either via the current direction, or anisotropic
pressure conditions.

Cell \#2 was designed to test the effect of anisotropic stress, in
analogy to similar work on \cps\ \cite{Demuer02}. The two samples
were aligned at right angles with respect to the cell axis. We
might expect the slight uniaxial stress associated with the solid
pressure medium to be oriented along this axis. The
superconducting and normal state properties did indeed differ
between the two samples in a way which corresponded to more than
just a shift of the pressure scale, which might be expected if
each sample was simply sampling part of an inhomogeneous pressure
distribution.
\begin{figure}
\begin{center}
\includegraphics[width=0.6\columnwidth]{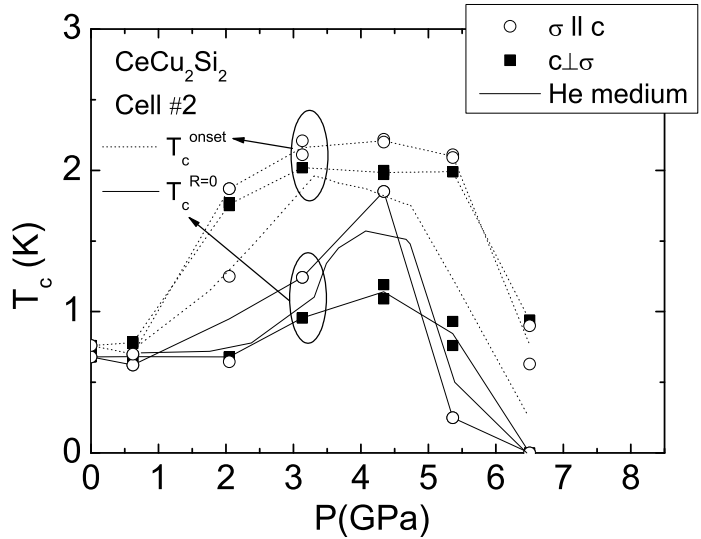}
\end{center}
\caption{\label{fig:straintc} Onset \tconset (dotted lines), and
completion $T_c^{R=0}$ (continuous lines) temperatures in cell \#2
compared with helium cell results. Two results are shown for
samples \sllc\ (open circles) and \sprc\ (filled squares), taken
from measurements using different contacts on the same sample. The
differences in \tc\ between samples are significantly larger than
local variations within a single sample.}
\end{figure}
Figure~\ref{fig:straintc} shows the onset \tconset, and completion
$T_c^{R=0}$ temperatures of the resistive superconducting
transition for samples \sllc\ and \sprc. $T_c^{R=0}$ has been
previously identified with the bulk superconducting transition
\cite{Holmes04a}. It is clear that the extension of \tconset\
above 6$\:$GPa seen in cell \#1 was not reproduced, indeed both
samples in cell \#2 follow the pressure dependence of `A/S
$I\parallel a$' fairly closely. However, the difference in $T_c$
between the two samples is an order of magnitude larger than
within each. There is a slight enhancement of $T_c$ in sample
\sllc, and a very clear difference in the shape of the resistive
transitions (see \cite{HolmesPhD} for details). The lack of
extension of the superconducting region to very high pressure in
cell \#2 does not entirely rule out uniaxial stress as the
explanation for such behaviour in cell \#1; the gasket instability
in the latter probably lead to higher non-hydrostatic stresses
inside the pressure cell.

The normal state properties provide a much clearer picture of the
effect of uniaxial stress on \ce\ under pressure. This can be most
simply described as a shift of the valence instability $P_{\rm v}$
to higher pressure in sample \sprc\ than in \sllc. We have
previously identified \pv\ with several anomalies in the normal
state resistivity, including:
\begin{itemize}
    \item Merging of the crystal field split Kondo resistivity
    maxima $T_1^{\rm max}$ and $T_2^{\rm max}$.
    \item A maximum in the residual resistivity $\rho_0$.
    \item A sudden change in the $A$ coefficient of the
    resistivity $\rho=\rho_0+AT^2$, and a change of the $A
    \propto(T_1^{\rm{max}})^{-2}$ scaling.
\end{itemize}

\begin{figure}
\begin{center}
\includegraphics[width=0.6\columnwidth]{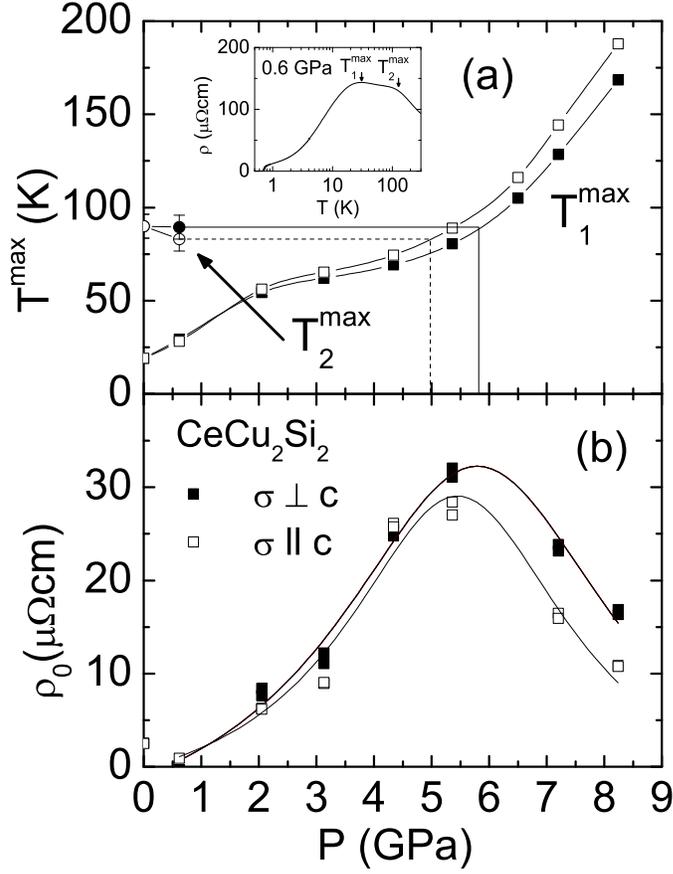}
\end{center}
\caption{\label{fig:tmaxrho0} (a) The crystal field split Kondo
resistivity maxima $T_1^{\rm max}$ and $T_2^{\rm max}$ (defined in
inset) merge at a higher pressure in sample \sprc\ (filled
symbols) than in \sllc\ (empty symbols). The uncertainty on
$T_1^{\rm max}$ is smaller than the symbol size. (b) The residual
resistivity maxima in the two samples corresponds approximately to
the pressures determined in (a). \pv\ is therefore at a higher
pressure in \sprc\ than in \sllc. The same symbols at a given
pressure correspond to measurements from different contacts on the
same sample.}
\end{figure}

\begin{figure}
\begin{center}
\includegraphics[width=0.6\columnwidth]{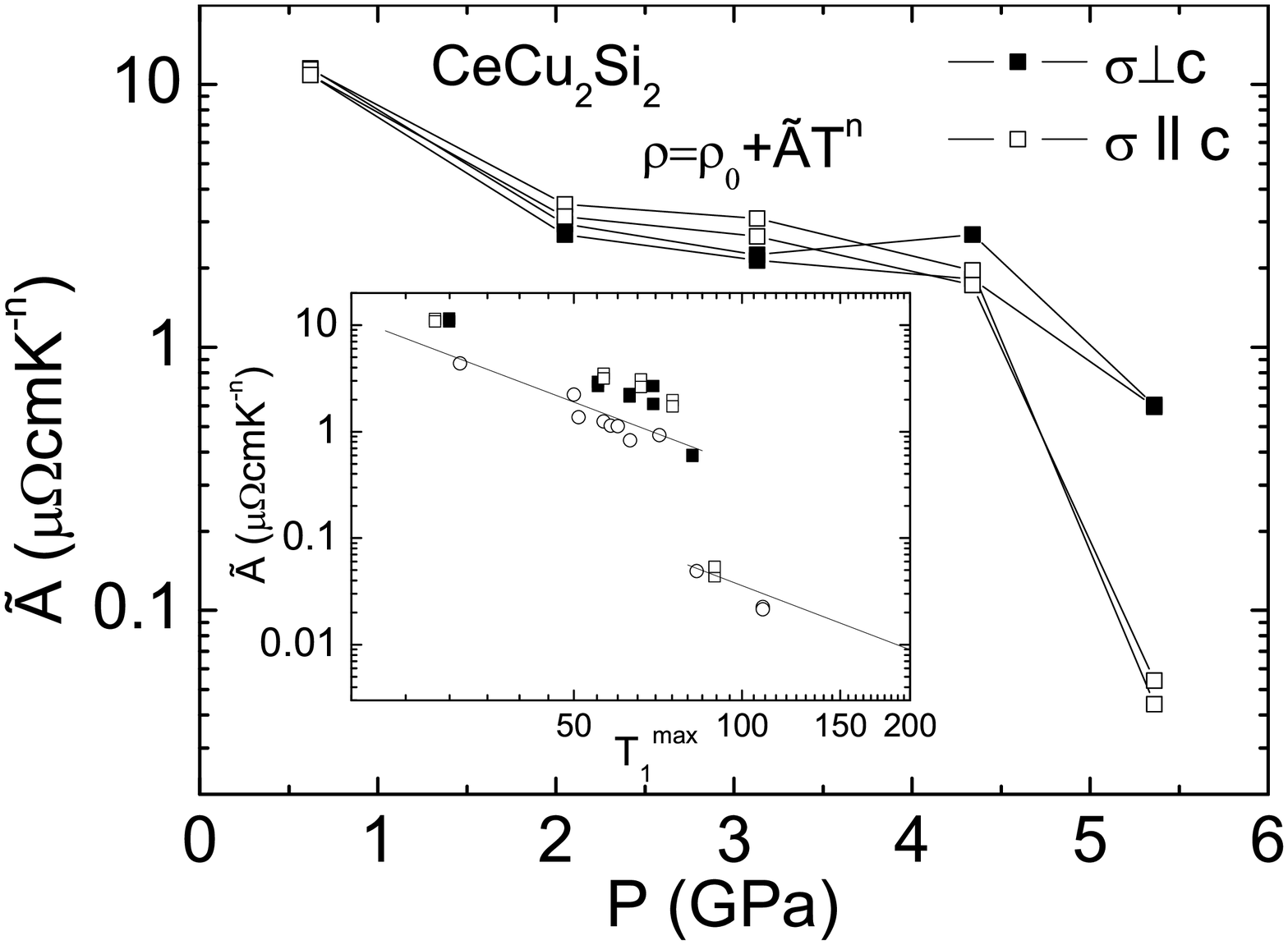}
\end{center}
\caption{\label{fig:strainAvsTmax} There is a drop in $\tilde{A}$
of more than an order of magnitude at the valence transition, at
different pressures for each sample. The inset shows the
corresponding change of the $A \propto(T_1^{\rm{max}})^{-2}$
scaling expected due to the Kadowaki-Woods relation (solid lines).
He cell results (open circles) using a quadratic fit are included
in the inset.}
\end{figure}

The shift of \pv\ is evident in all three properties, as shown in
Figs.~\ref{fig:tmaxrho0} \& \ref{fig:strainAvsTmax}. In
Fig.~\ref{fig:tmaxrho0}(a), one can see that $T_1^{\rm max}$,
proportional to the Kondo temperature $T_K$, rises faster in
sample \sllc\ than in \sprc. $T_2^{\rm max}$, which reflects the
crystal field splitting $\Delta_{\rm CEF}$, is difficult to
distinguish above the lowest pressure, but it has been shown to
remain more or less constant \cite{Jaccard99}. The two
crystal-field split resistivity peaks merge into a single maximum
at different pressures in the two samples, and this corresponds
approximately to the $\rho_0$ maximum. The latter is shown in
Fig.~\ref{fig:tmaxrho0}(b), where a lorentzian peak can be fitted
to the variation of $\rho_0$. Note that the values of $\rho_0$
plotted in Fig.~\ref{fig:tmaxrho0}(b) are extracted from a power
law fit to the resistivity above $T_c$, so the very low values at
0.6$\:$GPa may be slightly anomalous. The presence of the A phase
at ambient pressure also makes direct comparison difficult.

The shift of \pv\ determined above corresponds to a difference of
around 0.5$\:$GPa between \sprc\ and \sllc. This is confirmed when
the $A$ coefficient of the resistivity is examined
(Fig.~\ref{fig:strainAvsTmax}). The analysis is slightly
complicated by the non-Fermi liquid behaviour of the resistivity
found in the entire region around $P_c$ and \pv, where exponents
$1<n<2$ are found in power law resistivity fits
$\rho=\rho_0+\tilde{A}T^n$. However in this case a reasonable
comparison is still possible on a logarithmic scale, since there
is a drop in $\tilde{A}$ of over an order of magnitude at \pv, and
even without forcing a quadratic fit, the variations in $n$ can
only contribute a maximum factor of two to the result. At higher
pressure (above 6$\:$GPa), the temperature dependent impurity
scattering dominated the resistivity and made it impossible to
obtain a reliable fit. The inset of Fig.~\ref{fig:strainAvsTmax},
where $\tilde{A}$ is plotted against $T_1^{\rm max}$, indicates
the crossover from the strongly to weakly correlated branches of
the Kadowaki-Woods plot \cite{Kadowaki86,MMV}. A comparison with
the helium cell results from Ref.~\cite{Holmes04a} (where a
quadratic fit was forced) shows that the drop in $\tilde{A}$
around 5$\:$GPa corresponds to the transition between the two
regions where the expected $A \propto(T_1^{\rm{max}})^{-2}$
scaling is followed (indicated by solid lines). The result is
clear.  The two samples track each other up to 4.5$\:$GPa, after
which the drop in $\tilde{A}$ occurs more quickly in \sllc\ than
\sprc, confirming that \sllc\ has reached the valence transition
at a lower pressure.

\section{Discussion}

Ishikawa attributed the differences at ambient pressure between
`low \tc ' and `high \tc ' type \ce\ to different pairing
symmetries \cite{Ishikawa03}. Normal state and superconducting
properties at ambient pressure differed considerably between the
two categories of sample, which were located in regions side by
side the homogeneity range \cite{Ishikawa01,Ishikawa02}. The `low
\tc ' samples were characterized by a slight excess of copper, and
a higher residual resistivity. However, specific heat measurements
showed a more robust superconducting transition in the `low \tc '
than the `high \tc ' samples. Despite the difference in ${\partial
T_c}/{\partial P}$ previously observed close to $P=0$, the `low
\tc ' and `high \tc ' samples both displayed the usual enhancement
of \tc\ at high pressure. The most striking difference was seen in
the residual resistivity, which reached huge levels in sample \#50
around \pv. Given that both samples had very similar but not
identical compositions, we might speculate on the origin of this.
According to Refs. \cite{Ishikawa03,Ishikawa01,Ishikawa02}, there
appeared to be no significant difference in crystal structure
between the two categories of sample. The exact location of any
atomic disorder may be highly significant, as the enhancement of
impurity scattering is due to critical fluctuations
\cite{Miyake02,Miyake02b} and the effect may well depend
significantly on the nature and location of the
impurity itself, not only at \pv, but also $P_c$. 
The effect of disorder on \tc\ noted by Yuan et
al.~\cite{Yuan03b}, suggests that the initial reduction in \tc\
with pressure in the `low \tc ' samples is due to such (possibly
enhanced) impurity scattering, suppressing superconductivity
between the two critical points.

The A/S type \ce\ shows a similar enhancement of \tc\ at high
pressure to the sample measured in hydrostatic conditions. All
other samples, including those of other types, have shown the
enhancement of $T_c$ when pressurized to over 3$\:$GPa. Regarding
both the properties at ambient pressure and around \pv, the
evidence so far seems to suggest only that the
A$\rightarrow$A/S$\rightarrow$S series represents a slight shift
of the pressure scale. Thus any individual \ce\ sample can be
classified by two or maybe three variables, corresponding to the
aforementioned pressure shift, and to the impurity concentration
(and perhaps also to the nature of the disorder).

Regarding the effect of uniaxial stress on \ce, Monthoux and
Lonzarich have shown that a more anisotropic structure leads to an
increase in \tc\ for both magnetic and density fluctuation
mediated pairing \cite{Monthoux04}. In the Ce115 systems, and
related Pu compounds, it has also been shown that \tc\ is strongly
dependent on the ratio of the tetragonal lattice parameters $c/a$
\cite{bauer:147005}. While the exact effect of the
quasihydrostatic medium is hard to quantify, our observations are
not inconsistent with these scenarios. This does not necessarily
help to distinguish between spin and valence-fluctuation mediated
superconductivity, but it provides a constraint which must be
satisfied by any more complete theoretical model. Further avenues
for exploration from both a theoretical and experimental point of
view come from the highly anisotropic resistivity, including some
evidence that \tc, as defined by zero resistivity, may depend on
current direction! One clear result not so far mentioned is that
the height of the resistivity peak at $T_1^{\rm max}$ is
significantly larger for $I\parallel a$ than $I\parallel c$.  This
is not the case for the magnitude of $\rho(T_2^{\rm max})$,
reflecting the higher symmetry of the higher-lying crystal field
split $f$-states. The effect of the CEF on the anisotropy of the
resistivity in \ce was analysed in Ref.~\cite{Kashiba86}.

Finally, we would like to emphasize again the importance of the
coexistence of superconductivity and very high residual
resistivity shown in Fig.~\ref{fig:5057p4}. This is perhaps the
most compelling evidence so far for a novel pairing mechanism in
\ce\ at high pressure.

\section{Conclusions}
Resistivity measurements were carried out at high pressure on
several different \ce\ samples, in the presence of small
non-hydrostatic stress components. All samples showed an
enhancement of superconductivity around 2--3$\:$GPa, and also an
enhancement of residual resistivity with a maximum around
5$\:$GPa. Complete resistive superconducting transitions can
coexist with residual resistivities of the order of
150--200$\:\mu\Omega$cm around the valence instability close to
5$\:$GPa. Individual \ce\ samples, usually classified at ambient
pressure by the presence or otherwise of an ordered `A' phase
and/or superconductivity, belong to the same continuum, which can
be traversed by pressure and/or Ge substitution.  Disorder can
also be used to classify samples, and the enhancement of residual
resistivity under pressure probes this. It may be a fruitful
avenue for future research to examine the effect of changing the
nature of the impurities. Uniaxial stress was found to have a
significant effect on both the normal and superconducting
properties around the valence instability, which can be summarized
by a shift of the valence instability pressure \pv.

\section*{Acknowledgements}
ATH wishes to thank the Japan Society for the Promotion of Science
for financial support during part of the preparation of this
manuscript. MI thanks Dr. P. Ahmet, now at National Institute for
Materials Science for his help in preparing samples by a
levitation method.


\section*{References}
\begin{thebibliography}{10}

\bibitem{Miyake99}
Miyake K, Narikiyo O, Onishi Y 1999 {\em Physica B} {\bf 259--261}
676--7

\bibitem{Onishi00}
Onishi Y, Miyake K 2000 {\em J. Phys. Soc. Jpn.} {\bf 69} 3955--64

\bibitem{Holmes04a}
Holmes A T, Jaccard D, Miyake K 2004 {\em Phys. Rev. B} {\bf 69}
024508

\bibitem{Yuan03b}
Yuan H, Grosche F M, Deppe M, Geibel C, Sparn G, Steglich F 2003
{\em Science}
  {\bf 302} 2104

\bibitem{Jaccard04}
Jaccard D, Holmes A T 2004 {\em Physica B} {\bf 359--361} 333--40

\bibitem{Gegenwart98}
Gegenwart P, Langhammer C, Geibel C, Helfrich R, Lang M, Sparn G,
Steglich F,
  Horn R, Donnevert L, Link A, Assmus W 1998 {\em Phys. Rev. Lett.} {\bf 81}
  1501--4

\bibitem{Trovarelli97}
Trovarelli O, Weiden M, Muller-Reisener R, Gomez-Berisso M,
Gegenwart P, Deppe
  M, Geibel C, Sereni J G, Steglich F 1997 {\em Phys. Rev. B} {\bf 56} 678--85

\bibitem{Miyake02}
Miyake K, Maebashi H 2002 {\em J. Phys. Soc. Jpn.} {\bf 71}
1007--10

\bibitem{Ishikawa83}
Ishikawa M, Braun H F, Jorda J L 1983 {\em Phys. Rev. B} {\bf 27}
3092--5

\bibitem{Modler95}
Modler R, Lang M, Geibel C, Schank C, M{\"u}ller-Reisener R,
Hellmann P, Link
  A, Sparn G, Assmus W, Steglich F 1995 {\em Physica B} {\bf 206--207} 586--8

\bibitem{Steglich96}
Steglich F, Gegenwart P, Geibel C, Helfrich R, Hellmann P, Lang M,
Link A,
  Modler R, Sparn G, Buttgen N, Loidl A 1996 {\em Physica B} {\bf 223--224}
  1--8

\bibitem{Louca00}
Louca D, Thompson J D, Lawrence J M, Movshovich R, Petrovic C,
Sarrao J L, Kwei
  G H 2000 {\em Phys. Rev. B} {\bf 61} R14940

\bibitem{Ishikawa03}
Ishikawa M, Takeda N, Koeda M, Hedo M, Uwatoko Y 2003 {\em Phys.
Rev. B} {\bf
  68} 024522

\bibitem{Demuer02}
Demuer A, Holmes A T, Jaccard D 2002 {\em J. Phys: Condens.
Matter} {\bf 14}
  L529--35

\bibitem{Jeevanpc}
Jeevan H S
\newblock Private communication

\bibitem{Jaccard98}
Jaccard D, Vargoz E, Alami-Yadri K, Wilhelm H 1998 {\em Rev. High
Pressure Sci.
  Techno.} {\bf 7} 412--18

\bibitem{HolmesPhD}
Holmes A T  2004
\newblock {\em Ph.D. thesis} University of Geneva

Available from:
http://www.unige.ch/cyberdocuments/theses2004/HolmesAT/these.pdf

\bibitem{Jaccard99}
Jaccard D, Wilhelm H, Alami-Yadri K, Vargoz E 1999 {\em Physica B}
{\bf
  259--261} 1--7

\bibitem{Kadowaki86}
Kadowaki K, Woods S 1986 {\em Solid State Commun.} {\bf 58} 507

\bibitem{MMV}
Miyake K, Matsuura T, Varma C M 1989 {\em Solid State Commun.}
{\bf 71} 1149

\bibitem{Ishikawa01}
Ishikawa M, Takeda N, Ahmet P, Karaki Y, Ishimoto H 2001 {\em J.
Phys: Condens.
  Matter} {\bf 13} L25--31

\bibitem{Ishikawa02}
Ishikawa M, Takeda N, Ahmet P, Karaki Y, Ishimoto H, Huo D,
Sakurai J 2002 {\em
  J. Phys. Chem. Solids} {\bf 63} 1165

\bibitem{Miyake02b}
Miyake K, Narikiyo O 2002 {\em J. Phys. Soc. Jpn.} {\bf 71}
867--71

\bibitem{Monthoux04}
Monthoux P, Lonzarich G G 2004 {\em Phys. Rev. B} {\bf 69} 064517

\bibitem{bauer:147005}
Bauer E D, Thompson J D, Sarrao J L, Morales L A, Wastin F,
Rebizant J, Griveau
  J C, Javorsky P, Boulet P, Colineau E, Lander G H, Stewart G R 2004 {\em
  Physical Review Letters} {\bf 93} 147005

\bibitem{Kashiba86}
Kashiba S, Maekawa S, Takahashi S, Tachiki M 1986 {\em J. Phys.
Soc. Jpn.} {\bf
  55} 1341

\end{thebibliography}
\end{document}